# Accelerating *ab initio* path integral molecular dynamics with multilevel sampling of potential surface

Hua Y. Geng[1,2]

[1] *National Key Laboratory of Shock Wave and Detonation Physics, Institute of Fluid Physics, CAEP; P.O.Box 919-102 Mianyang, Sichuan, P. R. China, 621900*

[2] *Department of Chemistry and Chemical Biology, Cornell University, Baker Laboratory, Ithaca, New York 14853, USA*

## Abstract

A multilevel approach to sample the potential energy surface in a path integral formalism is proposed. The purpose is to reduce the required number of *ab initio* evaluations of energy and forces in *ab initio* path integral molecular dynamics (AI-PIMD) simulation, without compromising the overall accuracy. To validate the method, the internal energy and free energy of an Einstein crystal are calculated and compared with the analytical solutions. As a preliminary application, we assess the performance of the method in a realistic model—the FCC phase of dense atomic hydrogen, in which the calculated result shows that the acceleration rate is about 3 to 4-fold for a two-level implementation, and can be increased to 10 times if extrapolation is used. With only 16 beads used for the *ab initio* potential sampling, this method gives a well converged internal energy. The residual error in pressure is just about 3 GPa, whereas it is about 20 GPa for a plain AI-PIMD calculation with the same number of beads. The vibrational free energy of the FCC phase of dense hydrogen at 300 K is also calculated with an AI-PIMD thermodynamic integration method, which gives a result of about 0.51 eV/proton at a density of $r_s = 0.912$.

## I. INTRODUCTION

The imaginary-time path integral provides an elegant and powerful formalism for studying the thermodynamic properties of many-body quantum systems.[1,2] By mapping a quantum particle onto an isomorphic classical polymer in which replicas (or *beads*) are connected via harmonic springs,[3] one avoids the cumbersome requirement of solving the Schrödinger equation for the wave functions. Many simulation techniques developed for classical system can then be applied to quantum systems directly.[4-6]

Briefly, in quantum statistical mechanics, if we let $H = \hat{T} + \hat{V}$ denote the Hamiltonian of the system, $\beta = (k_B T)^{-1}$ be the inverse of the temperature, then the canonical quantum partition function $Z(\beta) = \text{Tr}\exp(-\beta H)$. Using Trotter's theorem[7] for the canonical density operator

$$e^{-\beta H} = \lim_{P\to\infty}\left(e^{-\left(\frac{\beta}{2P}\right)\hat{V}}e^{-\frac{\beta}{P}\hat{T}}e^{-\left(\frac{\beta}{2P}\right)\hat{V}}\right)^P, \qquad (1)$$

the partition function can be rewritten as a path integral

$$Z(\beta) = \text{Tr}\left(e^{-\frac{\beta}{P}H}\right)^P = \int d\mathbf{r}_1 \cdots \int d\mathbf{r}_P\, \rho(\mathbf{r}_1,\cdots,\mathbf{r}_P;\beta), \qquad (2)$$

where the density matrix $\rho(\mathbf{r}_1,\cdots,\mathbf{r}_P;\beta) = \prod_{j=1}^{P}\rho(\mathbf{r}_j,\mathbf{r}_{j+1};\beta/P)$, and

$$\rho(\mathbf{r}_1,\cdots,\mathbf{r}_P;\beta) \propto \exp\left(-\sum_{j=1}^{P}\frac{mP}{2\beta\hbar^2}(\mathbf{r}_{j+1}-\mathbf{r}_j)^2 - \frac{\beta}{P}\sum_{j=1}^{P}V(\mathbf{r}_j)\right), \qquad (3)$$

in which $\mathbf{r}_j$ is the system coordinate at the $j$th time slice (or bead) with the cyclic condition $\mathbf{r}_0 = \mathbf{r}_P$. When $P$ takes a finite value, the primitive approximation is obtained. The density matrix in this form of path integral can be sampled using Monte Carlo (PIMC)[2,4] or molecular dynamics (PIMD)[5,6] methods, in which any observables of a quantum system in an *NVT* ensemble can be obtained via an ensemble average

$$\langle A \rangle = \langle A \rangle_{NVT} = \frac{1}{Z(\beta)}\text{Tr}\left[A\exp(-\beta H)\right]. \qquad (4)$$

In most applications of the path integral method in chemistry and condensed matter physics, the potential energy $V(\mathbf{r})$ is given in terms of inter-atomic interaction potentials. The potential function may be explicit or implicit. For the former the potential must be defined in advance, whereas it can be generated on-the-fly in the latter case, usually by *ab initio* methods such as density functional theory (DFT), and the method dubbed *ab initio* path integral method (AI-PI).[8-11]

Both implementations have their respective merits and demerits. Algorithms employing explicit potentials are usually much faster in computation. But an accurate potential is difficult to obtain, and might be subject to transferability problem,[12] especially under high pressure conditions. On the other hand, though an *ab initio* potential can be high in accuracy and, in principle, without any transferability issue, the computational demand is huge. Therefore it will be beneficial if one can combine the merits of the two approaches. Namely, exploit explicitly predefined inter-atomic potential to reduce the total computational cost in the AI-PI method, but without sacrificing the overall accuracy.

From the construction of the path integral formalism as shown above, it is evident that the computational cost mainly comes from two sources: (*i*) evaluating the energy and forces of a single bead, and (*ii*) repeating the same process for all beads. Therefore, the computational cost will be diminished if one can reduce the required total number of beads. At low temperatures or for light elements, however, employing fewer beads usually implies a bad short-time propagator for the density matrix.[2] There have some techniques been developed to improve the short-time propagators so that a small number of beads can be used, such as the pairwise action approximation[2,13] and high-order composite factorizations in Eq.(1).[14-19] Unfortunately, these require either a predefined interaction potential[2,13] and/or second or higher order potential derivatives,[14-18] or having negative coefficients,[19] and thus cannot be implemented in AI-PIMD directly.[18] A completely different approach, the ring polymer contraction (RPC) scheme proposed by Markland and Manolopoulos, shows great promise in this respect.[20,21] The challenge along this line is how to split and arrange the forces into short-range and long-range contributions in *ab initio* simulations.[22] We shall describe how to remove this difficulty in this paper by introducing an arbitrary splitting of a potential into additive parts, using simple model reference potential as a demonstration. Formally, our concept is equivalent to improving the short-time propagator by using a multi-level sampling technique with predefined approximate potentials. In following discussions we will ignore the cost of evaluating the predefined approximate inter-atomic potential, since it is tiny by comparison with that for an *ab initio* potential. We discuss PIMD only, the extension to PIMC is straightforward.

Another method that is very similar, but not identical, to RPC is the mixed quantum-classical scheme and its improved version of mixed time slicing (MTS) procedure.[23,24] MTS is intended to optimize the quantization of different degrees of freedom (*e.g.*, those of the light and heavy particles) using different number of beads,[24] whereas RPC is purposed to accelerate the calculation by sampling the components of the potential that have different spatial variation with different number of beads, in which the procedure usually is carried out to all degrees of freedom in a parallel fashion.[20] The spirit and implementation of these two methods are not the same, but closely related. Though we will not discuss MTS in detail in this paper, an efficient combination of the two, which as an interesting extension of the proposed multi-level sampling scheme and a unification of RPC and MTS, will be given at the end of the paper.

The paper is organized as follows: the theoretical basis and algorithm are presented in the next section, in which both the multilevel technique for potential energy surface sampling and its implementation in PIMD, as well as the thermodynamic integration using PIMD (TI-PIMD) to calculate the free energy are discussed. In Sec.III we discuss the application to an Einstein crystal, for which the analytical solution is known and thus serves as a validation of the approach. A preliminary application to a realistic system is given in Sec.III B, where we apply the method to the FCC phase of dense monatomic hydrogen under high pressure. Section IV concludes the paper with a discussion and a summary.

II. THEORY AND ALGORITHM

A. Path integral molecular dynamics

The partition function for a canonical ensemble of distinguishable particles in quantum statistical mechanics, if expressed in the path integral formalism, is[1,2]

$$Z = \left(\frac{mP}{2\pi\beta\hbar^2}\right)^{\frac{3NP}{2}} \int \exp\left(-\sum_{j=1}^{P}\frac{mP}{2\beta\hbar^2}(\mathbf{r}_{j+1}-\mathbf{r}_j)^2 - \frac{\beta}{P}\sum_{j=1}^{P}V(\mathbf{r}_j)\right)[d\mathbf{r}]^{NP}. \quad (5)$$

This expression is formally equivalent to that of classical ring polymers.[1-5] Namely, an $N$-particle quantum system in a potential of $V(\mathbf{r})$ at a temperature $\beta^{-1}$ can be mathematically mapped onto a classical system of ring polymers that have $NP$ particles interacting via an effective potential $\tilde{V}(\mathbf{r})$ at a temperature of $P\beta^{-1}$, where

$$\tilde{V}(\mathbf{r}) = \sum_{j=1}^{P}\frac{mP^2}{2\beta^2\hbar^2}(\mathbf{r}_{j+1}-\mathbf{r}_j)^2 + \sum_{j=1}^{P}V(\mathbf{r}_j). \quad (6)$$

This mapping is not unique. For a quantum canonical ensemble at a constant temperature, sampling of the density matrix is independent of the pre-factor in Eq.(5). Therefore one can choose an arbitrary (but positive) pre-factor for the purpose of facilitating phase-space exploration.[5,25] For example we can choose to sample the system at a temperature of $\beta^{-1}$; then the corresponding effective potential becomes

$$\tilde{V}(\mathbf{r}) = \sum_{j=1}^{P}\frac{mP}{2\beta^2\hbar^2}(\mathbf{r}_{j+1}-\mathbf{r}_j)^2 + \frac{1}{P}\sum_{j=1}^{P}V(\mathbf{r}_j). \quad (7)$$

The isomorphism guarantees that the classical motion equations generated from the Lagrangian

$$\mathcal{L} = \sum_{j=1}^{P}\sum_{I=1}^{N}\left(\frac{\mathbf{p}_{I,j}^2}{2m'} - \frac{mP}{2\beta^2\hbar^2}(\mathbf{r}_{I,j+1}-\mathbf{r}_{I,j})^2 - \frac{1}{P}V(\mathbf{r}_{I,j})\right) \quad (8)$$

reproduce the correct quantum statistics of a canonical ensemble. Here we have rewritten $\mathbf{r}$ as a $3NP$ dimensional vector and the fictitious mass $m'$ can take any positive value. The corresponding motion equations are[8]

$$m'\ddot{\mathbf{r}}_{I,j} = -\frac{mP}{\beta^2\hbar^2}(2\mathbf{r}_{I,j}-\mathbf{r}_{I,j+1}-\mathbf{r}_{I,j-1}) - \frac{1}{P}\frac{\partial V(\mathbf{r}_{I,j})}{\partial \mathbf{r}_{I,j}}. \quad (9)$$

Evolving the system state by integrating these equations iteratively, one finally reaches a canonical distribution of quantum states (in a path integral formalism), and therefore the consequent equilibrium quantum thermodynamics.[8,9,25]

It is obvious that the first term at the right hand side of Eq.(9) determines the quantization of the system. At high temperature and/or with heavy particles, the spatial size of the ring polymers that represent the integral path along imaginary time shrinks to a point, and thus diminishes the quantum dispersion. The number of beads $P$ has a similar function, which originates from the Trotter factorization error introduced in Eq.(1).[26,27] It is thus necessary to use a large enough $P$ in order to catch the full quantum feature when carrying out the discrete path integral simulations. A commonly used rule of thumb is to check whether the short-time propagator is a good approximation or not. Usually the higher the short-time temperature is, the better the approximation becomes. Therefore one should choose a value of $P$ to ensure $\tau = \beta/P$ is small enough so that at this *temperature* the system behaves classically.[2,26,27]

Another factor that received less attention is that the optimal value of $P$ also depends on the shape of the potential energy surface. According to Bohr's correspondence principle, classical physics emerges from quantum mechanics in the limit of large quantum numbers, where some kind of continuity comes to the energy spectrum of the system. This usually requires a smooth potential which varies over scales much larger than the wavelength; or equivalently, the energy level spacing should be less than $\beta^{-1}$ at a finite temperature. With this consideration, one can reduce the required value of $P$ if a smooth potential surface is there. A recent development with great importance in this direction is RPC,[20] which is based on the well-established result that the minimal number of beads required to converge a PI calculation is $P > \beta\hbar\omega_{max}$, where $\omega_{max}$ is the maximum frequency present in the problem. By supposing that the potential terms are broken up into intramolecular and intermolecular parts, RPC proceeds with transforming the ring polymer to the normal mode representation, discarding the normal modes with high frequencies, and then transforming back to the coordinate space to evaluate the intermolecular potential which now has reduced beads.[20] The implementations of this method in simulations with empirical force fields showed that this approach is quite encouraging.[20-22] Instead of using normal mode transformation, we will attempt below an alternative realization of RPC directly in the coordinate space by using a hierarchical multilevel arrangement of the beads. This reformulation allows us to combine it with the MTS scheme naturally, in addition to some numerical benefits. To go further, we will introduce an arbitrarily defined model potential into this representation. We will then show that RPC in this form in fact can be interpreted as an effective approach to improve the short-time propagators, like as the well-known pairwise action approximation.[2] This improvement eventually allows us to apply RPC method to *ab initio* simulations directly.

B. Multilevel sampling of potential surface

We start with the exact operator identity[2]

$$\exp\left(-\tau(\hat{T}+\hat{V}) + \frac{\tau^2}{2}[\hat{T},\hat{V}]\right) = \exp(-\tau\hat{T})\exp(-\tau\hat{V}). \qquad (10)$$

The primitive approximation is good when the term $\frac{\tau^2}{2}[\hat{T},\hat{V}]$ becomes insignificant. Since the kinetic energy operator $\hat{T} \sim \nabla^2$, the magnitude of this term is approximately determined by (*i*) the short-time step size $\tau$, and (*ii*) the Laplacian of the potential $\nabla^2\hat{V}$. If the potential is flat and has a small curvature, few beads can be used in the primitive approximation

$$\exp\left(-\tau(\hat{T} + \hat{V})\right) \approx \exp(-\tau\hat{T})\exp(-\tau\hat{V}). \qquad (11)$$

The minimum value of $P$ is constrained by the condition of $\tau|\nabla^2\hat{V}| \ll |\hat{T} + \hat{V}|$. The RPC, as well as the multilevel technique that we will propose below, exploit this property by decomposing the potential into several additive parts that have different spatial variations. For example, if a given potential can be separated into a rapidly varying "short-range" part and a slow-changing "long-range" part, and if the short-range part can be evaluated very fast whereas the long-range part is not, the above reasoning indicates that we can use different short-time steps (*i.e.*, different values of $P$, or equivalently, multiple imaginary time steps) to accelerate the whole calculation, without compromising the overall accuracy.

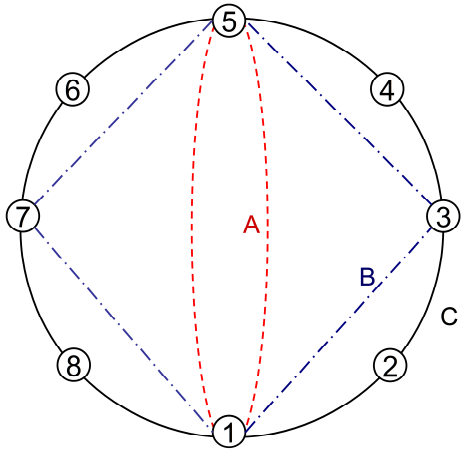

FIG. 1: (color online) Illustration of a multilevel sampling layout: The lowest level (*C*) contains all beads, and higher levels (*B* and *A*) contain only a subset of the lower levels.

Without using the normal mode transformation,[20] here we reformulate the problem directly in the coordinate space. To this end we introduce a multilevel arrangement of beads. Figure 1 illustrates the scheme of a multilevel sampling layout. The beads are constructed into hierarchical levels: the lowest one contains all beads, and higher levels are a subset of the lower levels. The beads at the highest level are called *primary beads*, and others are referred to as *extended beads*. The ratio of the bead numbers between contiguous levels is called *expansion factor*. For example in Fig.1, the lowest level *C* contains all beads that are indexed from 1 to 8; the next level *B* is a subset of *C* and contains only 4 beads (1, 3, 5, and 7). The highest one (level *A*) is a subset of both *B* and *C*, and contains only bead 1 and 5; they compose the primary beads. The expansion factor for this three-level layout is 4:2:1. Alternatively, if we take only level *A* and *C*, then the expansion factor of this two-level scheme is 4:1. This hierarchical arrangement allows for a separate sampling of different components in a potential by using different number of beads, thus reducing the total computational cost.

For clarity, hereinafter we focus only on a two-level scheme. The generalization to three or more levels is straightforward. An arbitrarily defined model potential $V_s(\mathbf{r})$ is introduced, which is for the purpose of splitting the potential into two components, one varies rapidly and the other is much smoother in the spatial domain, *i.e.*,

$$V(\mathbf{r}) = V_s(\mathbf{r}) + V_l(\mathbf{r}). \tag{12}$$

Obviously $V_s(\mathbf{r})$ and $V_l(\mathbf{r})$ here correspond to the intramolecular and intermolecular part in RPC,[20] respectively. In the same spirit of RPC, we wish to use a small $P$ to integrate the smooth part $V_l(\mathbf{r})$, which is supposed to be difficult to evaluate (*e.g.*, with *ab initio* methods), to reduce the computational demands. But more beads will be used for the rapidly varying but easy to evaluate part $V_s(\mathbf{r})$, to maintain the overall approximation accuracy. For this purpose, we rewrite Eq.(5) as

$$Z = \left(\frac{mP}{2\pi\beta\hbar^2}\right)^{\frac{3NP}{2}} \int \exp\left(-\sum_{j=1}^{P} \frac{mP}{2\beta\hbar^2}(\mathbf{r}_{j+1}-\mathbf{r}_j)^2 - \frac{\beta}{P}\sum_{j=1}^{P} V_s(\mathbf{r}_j) \right.$$
$$\left. - \frac{\beta}{L}\sum_{i=1}^{L}\sum_{j=1}^{P} V_l(\mathbf{r}_i)\tilde{\delta}_{ij}\right)[d\mathbf{r}]^{NP}, \tag{13}$$

where $\tilde{\delta}_{ij}$ is an abbreviated notation of $\delta_{Y_l(i),Y_s(j)}$, and takes a value of 1 when $Y_l(i)$ and $Y_s(j)$ *denote* the same bead, and equals 0 otherwise. It is worthwhile to note that since the set of beads $\{\mathbf{r}_i\}$ for the long-range part of the potential is a subset of all beads $\{\mathbf{r}_j\}$, and in general one can choose an equally distributed layout as demonstrated in Fig.1, the expression of the long-range part in Eq.(13) is actually equivalent to $-\frac{\beta}{P}\sum_{j=1}^{P}\overline{V}_l(\mathbf{r}_j)$, where $\overline{V}_l(\mathbf{r}_j) = V_l(\mathbf{r}_i)$ when $Y_s(j) = Y_l(i)$, and is a linear interpolation between $V_l(\mathbf{r}_{i+1})$ and $V_l(\mathbf{r}_i)$ when $Y_l(i) < Y_s(j) < Y_l(i+1)$, which is $\overline{V}_l(\mathbf{r}_j) = V_l(\mathbf{r}_i) + \frac{V_l(\mathbf{r}_{i+1})-V_l(\mathbf{r}_i)}{P/L}(Y_s(j) - Y_l(i))$. Here $Y_s(j) = j$ and $Y_l(i) = (i-1)\frac{P}{L} + 1$ are index functions of the beads according to the layout of the multilevel scheme as shown in Fig.1.

In this representation, whose appearance is more similar to MTS,[24] the underlying physical reasoning that why one can make the arbitrary splitting of the potential becomes evident. The short-time propagator for the density matrix, according to Eq.(13), now can be rewritten as

$$\rho\left(\mathbf{r}_i, \mathbf{r}_{i+1}, \frac{\beta}{L}\right) \propto \exp\left(-\frac{\beta}{2L}(V(\mathbf{r}_i) + V(\mathbf{r}_{i+1}))\right)$$
$$\times \int \exp\left(-\sum_{j}' \frac{mP}{2\beta\hbar^2}(\mathbf{r}_{j+1}-\mathbf{r}_j)^2 - \frac{\beta}{P}\sum_{j}'' V_s(\mathbf{r}_j)\right)[d\mathbf{r}]^{N(P/L-1)},$$

where $\Sigma'$ indicates a summation over the extended beads which satisfy $Y_l(i) \le Y_s(j) < Y_l(i+1)$, and $\Sigma''$ is a summation over $Y_l(i) < Y_s(j) < Y_l(i+1)$. It is now clear that the method expressed in this form can be viewed as an approach of improving the short-time propagator by integrating over an approximate potential $V_s(\mathbf{r})$. This justifies the arbitrary splitting of the potential in Eq.(12), by requiring that $V_s(\mathbf{r})$ is a good approximation to the true potential $V(\mathbf{r})$ in the spatial region that is accessible for particles within the given imaginary time step $\beta/L$.

The classical motion equations corresponding to Eq.(13) in this representation now become

$$m'\ddot{\mathbf{r}}_{I,j} = -\frac{mP}{\beta^2\hbar^2}\left(2\mathbf{r}_{I,j} - \mathbf{r}_{I,j+1} - \mathbf{r}_{I,j-1}\right) - \frac{1}{P}\frac{\partial V_s(\mathbf{r}_{I,j})}{\partial \mathbf{r}_{I,j}}$$
$$-\frac{1}{L}\sum_{i=1}^{L}\frac{\partial V_l(\mathbf{r}_{I,i})}{\partial \mathbf{r}_{I,i}}\tilde{\delta}_{ij}. \qquad (14)$$

Integrating these equations leads to a canonical ensemble of the distribution of quantum states. The energy is evaluated along the PIMD simulation using a thermodynamic estimator[4] of

$$E_{\text{tot}} = E_{\text{pot}} + E_{\text{kin}}, \qquad (15)$$

with the potential part

$$E_{\text{pot}} = \langle \frac{1}{P}\sum_{j=1}^{P} V_s(\mathbf{r}_j) + \frac{1}{L}\sum_{i=1}^{L}\sum_{j=1}^{P} V_l(\mathbf{r}_i)\tilde{\delta}_{ij} \rangle \qquad (16)$$

and the quantum kinetic energy of the particles

$$E_{\text{kin}} = \langle \frac{3NP}{2\beta} - \frac{mP}{2\beta^2\hbar^2}\sum_{j=1}^{P}\sum_{I=1}^{N}(\mathbf{r}_{I,j+1} - \mathbf{r}_{I,j})^2 \rangle, \qquad (17)$$

where $\langle \cdots \rangle$ denotes an ensemble average. Alternatively, one can also use the virial estimator[28,29] (for unbound systems)

$$E_{\text{kin}}^{\text{v}} = \langle -\frac{mP}{2\beta^2\hbar^2}\sum_{j=1}^{P}\sum_{I=1}^{N}(\mathbf{r}_{I,j+2} - \mathbf{r}_{I,j+1})\cdot(\mathbf{r}_{I,j+1} - \mathbf{r}_{I,j}) \rangle$$
$$+ \langle \frac{1}{4}\sum_{j=1}^{P}\sum_{I=1}^{N}(\mathbf{r}_{I,j+1} - \mathbf{r}_{I,j})$$
$$\cdot \left[\frac{1}{P}\left(\frac{\partial V_s(\mathbf{r}_{I,j+1})}{\partial \mathbf{r}_{I,j+1}} - \frac{\partial V_s(\mathbf{r}_{I,j})}{\partial \mathbf{r}_{I,j}}\right)\right. \qquad (18)$$
$$\left. + \frac{1}{L}\sum_{i=1}^{L}\frac{\partial V_l(\mathbf{r}_{I,i})}{\partial \mathbf{r}_{I,i}}(\tilde{\delta}_{i,j+1} - \tilde{\delta}_{i,j})\right] \rangle.$$

The pressure is given by the volume derivative of the free energy[2,28]

$$P_{ress} = -\left.\frac{dF}{d\Omega}\right|_\beta = \frac{1}{3\Omega}\langle 2E_{\text{kin}} - \sum_{j=1}^{P}\frac{\mathbf{r}_j \cdot \nabla V_s(\mathbf{r}_j)}{P} - \sum_{i=1}^{L}\sum_{j=1}^{P}\frac{\mathbf{r}_i \cdot \nabla V_l(\mathbf{r}_i)\tilde{\delta}_{ij}}{L}\rangle, \quad (19)$$

in which the definition of $e^{-\beta F} = Z$ has been used.

C. Thermodynamic integration

A benefit of the multilevel sampling method is that to calculate the quantum free energy with *ab initio* thermodynamic integration[30-33] might become easier, because of the less demanding AI-PI simulation for each discrete integration point. Let $V_i$ and $V_t$ denote the potential energy of the reference and the target system, respectively. Then one can construct a reversible path to connect these two systems. In a nearly linear thermodynamic integration scheme,[34-36] the potential energy at a point $\lambda$ ($\lambda$ is a coupling parameter. And as it varies from 0 to 1, the system is changed from the reference end to the target end) along the employed path can be expressed as

$$V(\lambda) = (1-\lambda)^k V_i + \lambda^k V_t. \quad (20)$$

Note that it becomes a linear coupling when $k = 1$. The reason to use the nearly linear scheme is because in a linear coupling the integrand diverges at the endpoints, a judicious choice of $k$ other than 1 may make the integrand behave well (in the following calculation we find $k = 2$ is suitable). Details about the performance of the nearly linear scheme can be found in Ref.[34-36]. Using the relation of $\beta F = -\ln Z$ and Eq.(5), we have the derivative of the free energy with respect to $\lambda$ as

$$\frac{\partial F(\lambda)}{\partial \lambda} = \langle \frac{1}{P}\sum_{j=1}^{P}\frac{\partial V(\mathbf{r}_j, \lambda)}{\partial \lambda}\rangle_\lambda. \quad (21)$$

Integrating this along the path we get the free energy of the target system as

$$F_t = F_i + \frac{k}{P}\int_0^1 \sum_{j=1}^{P}\left[\lambda^{k-1}\langle V_t(\mathbf{r}_j)\rangle_\lambda - (1-\lambda)^{k-1}\langle V_i(\mathbf{r}_j)\rangle_\lambda\right] d\lambda. \quad (22)$$

Here $\langle \cdots \rangle_\lambda$ corresponds to an ensemble average of the enclosed quantity at the point $\lambda$ along the path. $F_i$ is the already known free energy of the reference system. Note when $P = 1$, this formalism goes back to the classical thermodynamic integration.[30]

D. Implementation

The equations of motion Eq.(14) are integrated using a time-reversible area-preserving multiple time step and velocity Verlet algorithm.[37] The phase space density evolves under the influence of the symmetrically split discrete time propagator

$$e^{i\delta t L} \approx e^{i\left(\frac{\delta t}{2}\right)L_l} \left( e^{i\left(\frac{\delta t_s}{2}\right)L_s} \left(e^{i\delta t_k L_k}\right)^{M_k} e^{i\left(\frac{\delta t_s}{2}\right)L_s} \right)^{M_s} e^{i\left(\frac{\delta t}{2}\right)L_l}, \qquad (23)$$

where $L = L_k + L_s + L_l$ is the Liouville operator associated with the total classical ring-polymer Hamiltonian, in which $L_k$ refers to that of the quantum kinetic part, and the other two are for the "short-range" and "long-range" interactions, respectively. The largest time step satisfies $\delta t = M_s \delta t_s = M_s M_k \delta t_k$. In Cartesian coordinates, the Liouville operator $L$ for a system having $f$ degrees of freedom is defined as[37]

$$iL = \sum_{j=1}^{f} \left[ \dot{r}_j \frac{\partial}{\partial r_j} + F_j \frac{\partial}{\partial p_j} \right], \qquad (24)$$

where $\{r_j, p_j\}$ are the position and its conjugate momenta of the system, and $F_j$ the corresponding force on the *j*th degree of freedom. In below calculations we take $\delta t = 1.0$ fs, and $M_s = 8, M_k = 20$. The Andersen thermostat[38] is employed to achieve the canonical ensemble *NVT*.

Because of Eq.(13), the potential separation in Eq.(12) does not depend on the physical construction of the system; one can choose any scheme one likes. For computational efficiency, here we use a simple analytic pairwise potential to model $V_s(\mathbf{r})$, and the long-range part then is given by $V_l(\mathbf{r}) = V(\mathbf{r}) - V_s(\mathbf{r})$. To exploit the power of multilevel sampling technique fully, it is important to have $V_s(\mathbf{r})$ as close to $V(\mathbf{r})$ as possible. In principle, the more accurate $V_s(\mathbf{r})$ is, the fewer beads are required for $V_l(\mathbf{r})$ part. We recommend to construct $V_s(\mathbf{r})$ by a force-matching[39-42] fitting to *ab initio* forces calculated at conditions similar to that one wants to investigate. This strategy usually gives a smooth enough $V_l(\mathbf{r})$.

III. APPLICATION: RESULTS AND DISCUSSION

A. Einstein crystal

As a preliminary validation to the method, we apply the multilevel sampling approach to an Einstein crystal, in which each atom in the lattice is assumed as an independent 3D quantum harmonic oscillator, and all atoms oscillate with the same frequency. The analytical solution is available for this model, thus provides a good benchmark to examine the validation and performance of the method.

In order to implement the multilevel sampling in this model, we artificially separate the total harmonic potential $V(x) = \frac{1}{2}m\omega_0^2 x^2$ into two parts: $V_s(x) = \frac{1}{2}m\omega_1^2 x^2$ and $V_l(x) = \Delta V = V(x) - V_s(x)$. Without loss of generality, we arbitrarily take $\omega_1 = 0.8\omega_0$, thus the "smooth" part $V_l(x)$ is also a harmonic potential with a characteristic frequency of $0.6\omega_0$.

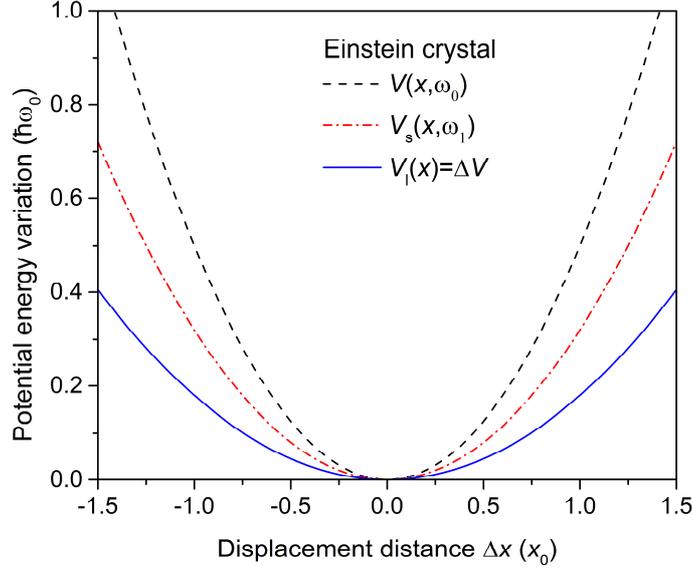

FIG. 2: (color online) Potential energy surfaces of an Einstein crystal as a function of the particle displacement distance in a unit of the natural length $x_0 = \sqrt{\hbar/m\omega_0}$.

Figure 2 illustrates the variation of the potential energy of an atom as it moves in an Einstein crystal. It can be seen that the separation of the harmonic potential into two parts reduces their spatial variation, as well as the curvature. For this case, the extra term $\frac{\tau^2}{2}[\hat{T},\hat{V}]$ in Eq.(10) is proportional to $\tau^2\omega^2$, thus the primitive approximation Eq.(11) will have the same level of accuracy for an imaginary time step $\tau_1 = \tau_0\,\omega_0^2/\omega_1^2$ in a harmonic potential characterized by $\omega_1$ as that of $\tau_0$ in a potential of $\omega_0$. That is to say, only a fraction (here ~0.36) of the originally demanded number of beads for $V(x)$ is required if to sample only the "smooth" part $V_l(x)$.

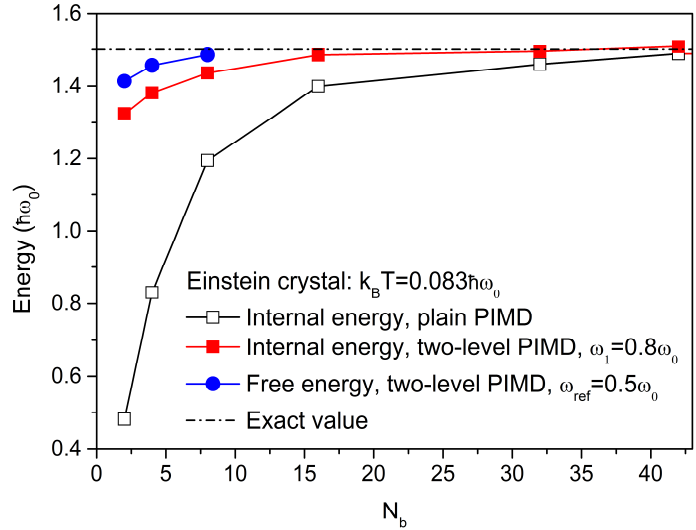

FIG. 3: (color online) Convergence of the calculated internal energy and free energy in an Einstein crystal with a vibration frequency of $\omega_0$ with respect to $N_b$, the number of (primary) beads employed in plain (two-level) PIMD simulations. The statistical error is smaller than the symbol size.

To demonstrate this, we calculate the canonical ensemble average of internal energy of the Einstein crystal by integrating the motion equations of a plain [Eq.(9)] and two-level scheme [Eq.(14)] PIMD to sample the potential energy surface. The simulation is performed at a temperature of $k_B T = 0.083 \hbar \omega_0$. In the plain PIMD simulations, the total number of beads to examine the convergence of the results with respect to the imaginary time step varies from 2 to 64. From Fig.3 we can see that the primitive approximation is good when the total number of beads is greater than 42 in a plain PIMD (by comparison with the dash-dotted line—the exact value).

In the case of a two-level multilevel sampling calculation, we set the total number of beads (that are employed to sample $V_s$) as 128, and change the number of the primary beads (those to sample $V_l$) from 2 to 42; namely, the expansion factor varies from about 64:1 to 3:1. Figure 3 clearly demonstrates the effect of the multilevel sampling scheme: in this two-level implementation, only 16 beads are required to sample the "smooth" part $V_l$. This gives an accuracy almost the same as the plain PIMD with 42 beads, *i.e.*, the required number of beads reduces to only a fraction of 0.38. This reduction rate is in line with the above simple analytical estimate.

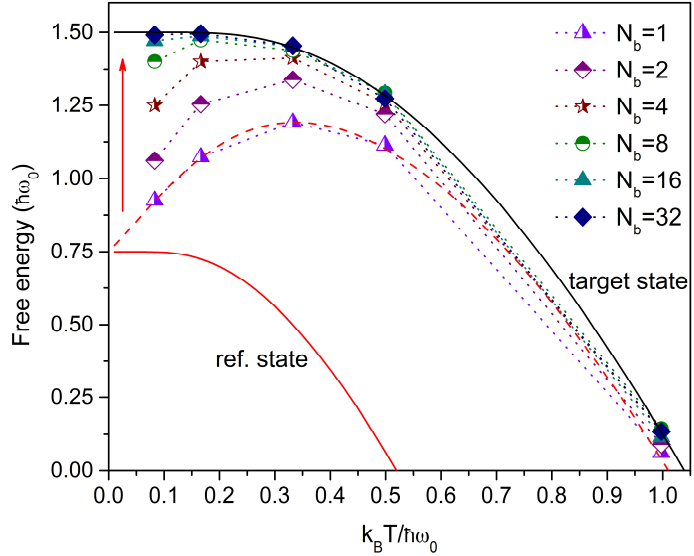

FIG. 4: (color online) Free energy of an Einstein crystal as a function of the temperature: solid lines—the exact value of the reference and target states, respectively; dashed line—the exact value of the (analytic) classical thermodynamic integration; points—this work, calculated with plain PIMD thermodynamic integration using different number of beads $N_b$; dotted lines are for guiding the eye only.

At this low temperature, the free energy is almost identical to the internal energy (the relative difference is less than 0.002%), therefore it is interesting to study how the free energy converges with respect to the number of primary beads. We first carry out a series of thermodynamic integration (TI) calculations with plain PIMD simulations, to examine the convergence behavior of the TI-PIMD with respect to the total number of beads. The reference state for TI calculation is chosen as an Einstein crystal with the oscillation frequency $\omega_r = 0.5\omega_0$. The results are shown in Fig.4, from which we can see that our TI-PIMD reproduces the exact free energy of the target system. As expected, at low temperatures a big value of $N_b$ is required. The classical TI (the special case with $N_b = 1$) can give reasonable results only when $k_B T \gg \hbar \omega_0$. Note that since this is a general feature of harmonic oscillators, the picture shown in Fig.4 also applies to the vibrational free energy in realistic materials such as molecules and solids.

When apply this TI-PIMD with the two-level sampling scheme, as Fig.3 shows, the calculated free energy converges to the exact value using only 8 primary beads. This is another twofold reduction in comparison with the internal energy case. That is, calculation of the free energy (as an integral quantity) allows using fewer beads in TI-PIMD. This feature can be traced back to the behavior as shown in Fig.4: the potential surface of the reference state is more flat than $V_l$, thus requires fewer $N_b$ to make Eq.(11) a good approximation along the TI path. Perceivable error is introduced only for those points near the target state, and has small impact on the final result. One can expect that if a reference state with steeper potential were employed, more $N_b$ would be required.

B. Realistic model: dense hydrogen

The above discussion of a simple theoretical model confirms that the multilevel sampling indeed can reduce the required number of primary beads in a path integral calculation. We will further demonstrate this by applying it with AI-PIMD simulations to a real material—dense hydrogen at high pressures, a system that has attracted much interest.[43,44] As an example, we discuss here only the monoatomic FCC phase of hydrogen, at a density characterized by $r_s = 0.912$, where the dimensionless parameter $r_s$ is defined as the radius of a sphere that encloses on the average one proton in a unit of the Bohr radius. This density corresponds to about 3.8 TPa in pressure, at a temperature of 300 K. At low pressures this high-symmetry phase is unstable.[45] But with increasing compression it gradually becomes stable. The lattice ordering is mainly determined by the combined effects of nuclear repulsion and mechanical compression, thus as a Wigner crystal of protons.[46]

In order to implement the multilevel sampling method, we divide the interatomic potential between hydrogen atoms into two parts: a steep pairwise potential which is described by an analytical function, and the slow-changing part which requires on-the-fly *ab initio* electronic total energy calculations.[47] To determine the pair potential, we employ the force matching method.[39-42] By using a $3 \times 3 \times 3$ supercell of FCC cubic structure containing 108 atoms, we sample a series of configurations from *ab initio* MD simulations equilibrated at 50K, 150K, 200K, 300K, 400K, and 550K, respectively. Then forces on all atoms in these configurations are computed. All of these calculations are performed with VASP code,[48] which is based on plane-wave basis and density functional theory of many-body electrons. The ion-electron interaction is described with all-electron like projector augmented-wave (PAW) potential,[49,50] and the Perdew-Burke-Ernzerhof (PBE) exchange-correlation functional[51] is used for the exchange-correlation functional. The Brillouin zone is sampled with a k-point grid of $2 \times 2 \times 2$, and the energy cut-off for the plane-wave basis set is taken as 600eV. The obtained forces are then fitted to a simple pair potential model $V_{PP}(r) = \alpha e^{-\varepsilon r}$. The result obtained is $\alpha = 23.36$ eV and $\varepsilon = 3.7$ Å$^{-1}$.

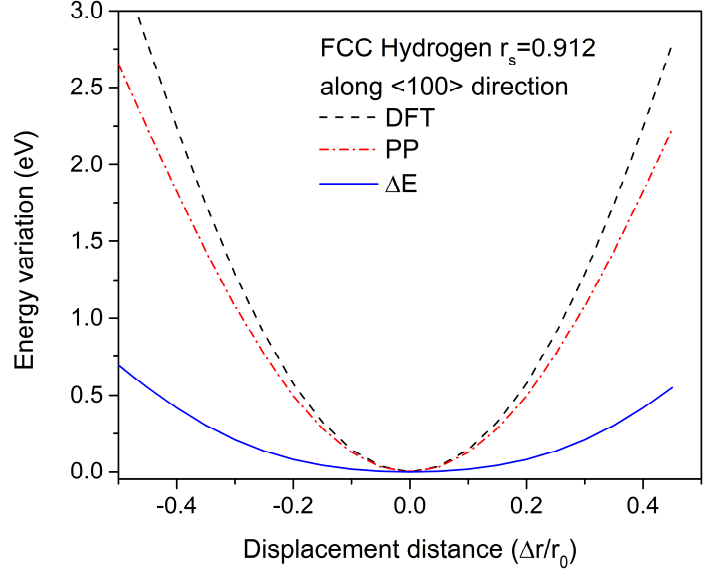

FIG. 5: (color online) Energy variation when one of the total 108 atoms in an FCC lattice of atomic hydrogen with $r_s$=0.912 is displaced along the <100> direction. Results calculated with DFT and pair potential are shown, as well as their difference $\Delta E$. Note $r_0$ is the nearest neighbor distance.

Figure 5 shows the potential energy variation when one H atom in the FCC supercell moves along the $\langle 100 \rangle$ direction. It can be seen that the simple repulsive pair-potential captures the DFT energy variation very well, so that the difference between them is very flat. This difference, $\Delta E(r) = E_{DFT}(r) - V_{PP}(r)$, is exactly the $V_l$ in Eq.(12) that is sampled with primary beads, whereas the "short-range" part is given by $V_s \equiv V_{PP}$. Since $V_{PP}$ is a simple analytical function, which can be evaluated very fast, using a large number of beads to sample this part do not increase the computational cost greatly. On the other hand, the most computationally demanding part—DFT total energy $E_{DFT}$—is calculated only when evaluating $\Delta E$. As shown in Fig.5, this part is much more flat, and the multilevel sampling technique would allow fewer beads for this part, which eventually will decrease the total required number of DFT calculations, and thus the overall computational cost.

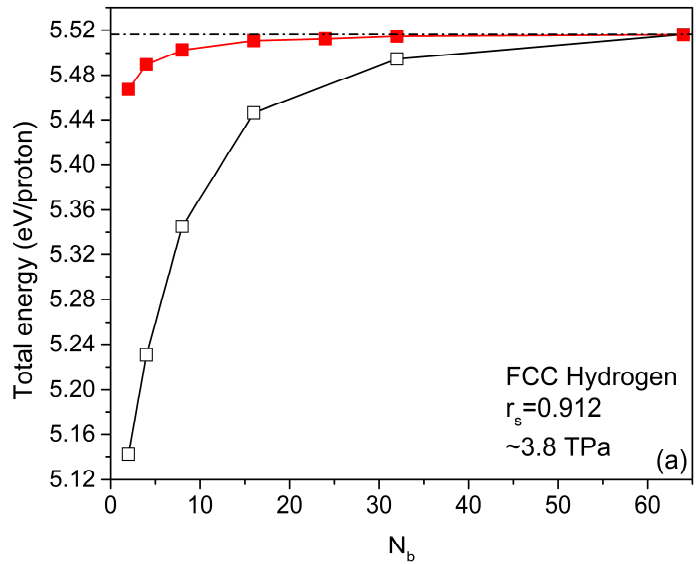

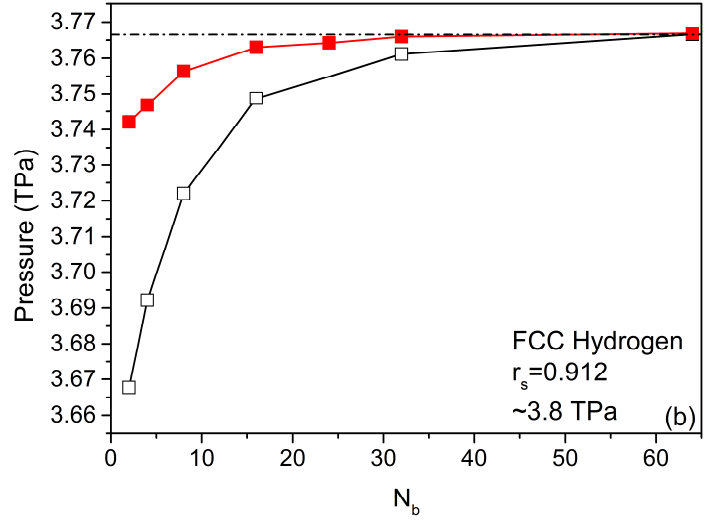

FIG. 6: (color online) Comparison of the (a) total energy and (b) pressure convergence with respect to the number of beads (or the primary beads) used in plain PIMD (open squares) and two-level PIMD (solid squares) simulations in FCC phase of atomic hydrogen at $r_s$=0.912. The data are obtained by averaging over a time scale of 2.0 ps, after a structural equilibration for about 1.0 ps.

Figure 6 illustrates the performance of the multilevel sampling method in dense hydrogen at 300K. In plain AI-PIMD simulations, the total energy converges with 64 beads. In comparison, for the two-level sampling method (for which 128 extended beads are used to sample the short-range model potential $V_{PP}$), within the statistical error, only 16 primary beads are required to give a converged result. That is to say, only one fourth of the *ab initio* runs are required in the latter case, a great reduction in the computational cost. The convergence in pressure is a little bit slower. Nevertheless the result from a multilevel sampling PIMD is always much better than a plain PIMD calculation. For example with 16 primary beads in the two-level AI-PIMD case, the error with respect to the converged value is just about 3 GPa. In comparison, with the same number of beads (thus the same burden in the *ab initio* calculation), the plain AI-PIMD predicted a pressure which is about 20 GPa lower. That is, using a two-level sampling can retrieve 85% of the error in pressure with this setting.

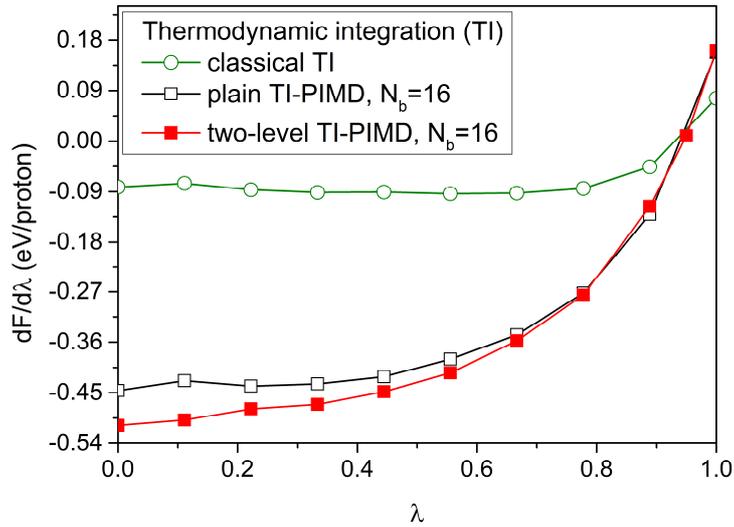

FIG. 7: (color online) Free energy derivative along the thermodynamic integration path, in which $\lambda$ is the coupling parameter. The simulation is performed at 300 K for FCC phase of atomic hydrogen with $r_s$=0.912. Each data point is obtained by averaging over a time scale of 2.0 ps, after a structural equilibration for about 1.0 ps.

The *ab initio* free energy of the same structure of dense hydrogen at the same thermodynamic condition as above is calculated using both classical and path integral TI. The target state corresponds to the coupling parameter $\lambda = 0$; and the reference state is an Einstein crystal at $\lambda = 1$, which has a zero point energy of 0.156 eV/proton at 0 K and a free energy of 0.154 eV/proton at 300 K. The integration is carried out with the nearly linear parameter $k = 2$.[34-36] As shown in Fig.7, the variation of the free energy derivative with this setting is smooth and close to be linear, except when approaching to the reference state, where the increment in $\frac{dF}{d\lambda}$ becomes steep. From the figure we can see that the classical TI fails; its free energy derivative is far from those of TI-PIMD. This is because dense hydrogen has a relatively high Debye temperature; along the TI path no classical MD simulation can capture the quantum feature of the lattice vibrations correctly. In contrast, plain TI-PIMD and two-level sampling TI-PIMD are close to each other. Only when one is near the target state, the plain TI-PIMD has a noticeable error. The free energy difference, from the thermodynamic integration of $\int_0^1 \frac{dF}{d\lambda} d\lambda$, is -0.074 eV/proton for classical TI, -0.334 eV/proton for plain TI-PIMD and -0.361 eV/proton for two-level TI-PIMD with 16 (primary) beads, respectively. As implied in Fig.6, the two-level TI-PIMD has converged at this condition, thus classical TI captures only about 21% of the total free energy difference, whereas plain TI-PIMD captures about 93%. The improvement of multilevel sampling here (recovers 7% of the total free energy difference) is not as striking as in the internal energy case, mainly because here we employed a reference state with a small characteristic frequency, thus a plain PIMD with 16 beads is already converged for those points near the reference state. If a reference state with higher frequency were used (which would make the curves shown in Fig.7 more linear, and thus a more accurate estimation of the final free energy), the improvement of the multilevel sampling would become more significant.

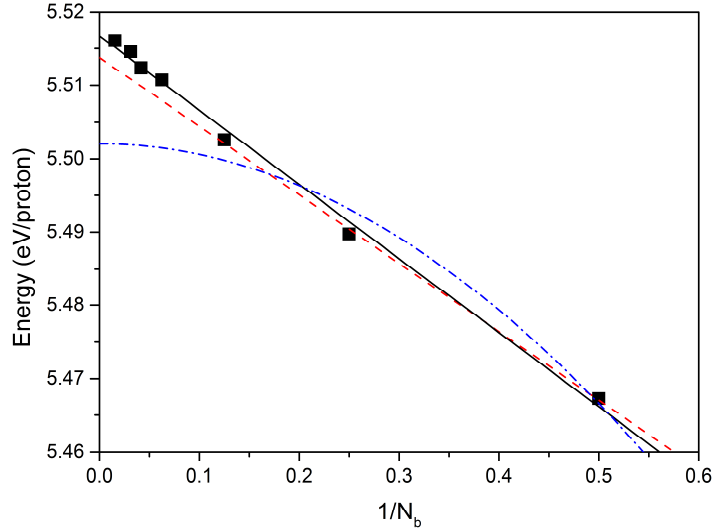

FIG. 8: (color online) Extrapolation behavior of the multi-level sampling scheme with respect to the number of primary beads. The solid line is a linear fitting to all simulation data (solid square points), and the dashed and dash-dotted lines are a linear and a quadratic fitting to the lowest three points ($N_b$=2, 4, and 8), respectively.

As a discretized path integral method as expressed in Eq.(13), one might tend to expect that it would converge asymptotically as $O(1/N_b^2)$. However, since it is not a standard Trotter decomposition, the actual behavior could be different. Instead of giving a mathematically rigorous proof, which is beyond the scope of the present paper, here we demonstrate the somewhat unexpected asymptotical scaling using a numerical example. Fig.8 plots the extrapolation behavior of the calculated energy with respect to the number of primary beads $N_b$, using the same simulation data as shown in Fig.6(a). It is evident from the figure that the asymptotic convergence is of $O(1/N_b)$. This linear convergence rate is slow when compared to the primitive approximation which goes as $O(1/N_b^2)$.[26] Therefore the acceleration gained in the multi-level sampling method mainly comes from the "short range" potentials that are accurate enough but easy to evaluate. On the other hand, since the convergence is slow, when very accurate results are required, making extrapolation becomes necessary. As shown in Fig.8, using extrapolation can further reduce the required number of *ab initio* evaluations more than two folds (to give the same level of energy estimate). Namely, with this method one can achieve an overall acceleration of about 8-10 times faster.

## IV. CONCLUDING REMARKS

In summary, the proposed multilevel sampling method indeed can reduce the computational cost of the *ab initio* part in AI-PIMD simulations. The acceleration rate is about 4 in dense hydrogen when a simple pairwise potential is used for the "short-range" part of the potential. If a more accurate model potential were used, it can be expected that the acceleration might be better. Using linear extrapolation can further reduce the computational cost more than two folds. This improvement for AI-PIMD is important, especially for light elements and their compounds at high pressures and low temperatures, where strong nuclear quantum effects usually require hundreds of beads in PIMD/PIMC simulation to get accurate results. It poses a huge computational burden if every bead requires an *ab initio* total energy calculation.

It is worthwhile to emphasize that by Eq.(13) our method is actually equivalent to improving the primitive short-time propagator by integrating over an approximate model potential [*i.e.*, the $V_s$ in Eq.(12)]. This feature, however, is obscured in the original formulation of RPC,[20] and thus makes it hard to justify the arbitrary splitting of the potential in that representation. The multilevel sampling algorithm, by its construction, is accurate up to only the first order.[52] The improvement comes with a very small error coefficient. Using traditional higher order corrections to the *ab initio* level in PIMD might be cumbersome, because of the requirement of higher order potential derivatives or the complexity of the algorithm.[14-19,53] Further reduction of the number of primary beads might be possible by using the generalized Langevin thermostat,[22,54,55] which is promising. Other schemes for asymptotic extrapolation other than the simple linear one as discussed above might also be helpful.[56-58] How much further improvement can be achieved by combining them still requires future investigations.

The concept of multilevel sampling is to shift the beads from *ab initio* level to the model potential level. Under some conditions, the total number of extended beads might become huge, and the computational cost on the model potential level might not be insignificant any more. In this situation, taking advantage of the fact that the potential is predefined, one can employ well-developed methods to reduce the total number of beads on this level, such as the pair-product action[2,13] or fourth-order method.[15,16,53]

Also note that the multilevel sampling is deliberately designed to accelerate AI-PI simulations, in which the potential separation is achieved by means of subtraction, which is arbitrary and independent of the actual construction of the system. This gives the multilevel sampling method a very large degree of flexibility. For example, one can easily insert an intermediate level to account for many-body effects by using predefined many-body model potentials, which would further reduce the number of primary beads required for *ab initio* calculations. An ideal implementation of the multilevel sampling is a four-level layout: using a simple pair potential at the first level, putting the many-body model potential at the second level, the DFT potential at the third level, and finally the most accurate quantum Monte Carlo (QMC) correction at the highest level. This arrangement would enhance our capability to achieve high theoretical accuracy while curbing the required computational cost.

Finally, it is helpful to combine RPC and MTS within the framework of the multi-level sampling scheme to give a unification formalism. The simplest MTS requires at least two degrees of freedom (DOF).[24] For a general two-dimensional system with DOFs $x$ and $y$, the unified expression thus can be written as

$$Z = \left(\frac{m_x P_x}{2\pi\beta\hbar^2}\right)^{\frac{P_x}{2}} \left(\frac{m_y P_y}{2\pi\beta\hbar^2}\right)^{\frac{P_y}{2}} \int [dx]^{P_x} \int [dy]^{P_y} \exp(-\beta V_f(x,y)), \qquad (25)$$

in which the effective potential $V_f = K + U$ is given by (see Eq.(23) in Ref.[24] for comparison)

$$K(x,y) = \frac{m_x P_x}{2\beta^2\hbar^2} \sum_{i=1}^{P_x} (x_{i+1} - x_i)^2 + \frac{m_y P_y}{2\beta^2\hbar^2} \sum_{i=1}^{P_y} (y_{i+1} - y_i)^2, \qquad (26)$$

and

$$U(x,y) = \frac{1}{P_x} \sum_{i=1}^{P_y} \sum_{j=1}^{P_x/P_y} V_s\left(x_{(i-1)\frac{P_x}{P_y}+j}, y_i\right) \\ + \frac{1}{L_x} \sum_{k=1}^{P_x} \sum_{i=1}^{L_y} \sum_{j=1}^{L_x/L_y} V_l(x_k, y_i) \tilde{\delta}_{k,(i-1)\frac{L_x}{L_y}+j}. \qquad (27)$$

Extension of it to more DOFs is straightforward. In this unified formalism, the quantization of different DOFs is optimized by MTS, and the calculation for each DOF is further accelerated with the RPC, thus benefits from the merits of both RPC and MTS schemes.

ACKNOWLEDGMENTS

The author would like to thank Professors R. Hoffmann and N. W. Ashcroft for their kind hospitality during his stay at Cornell University. Appreciation is also due Prof. R. Hoffmann and Dr. T. Zeng for their careful reading of the manuscript and for many pertinent comments. This work was supported by the National Natural Science Foundation of China under Grant No.11274281, the CAEP Research Project 2012A0101001, the National Science Foundation through Grant No. CHE-0910623, and also by EFree, an Energy Frontier Research Center funded by the U.S. Department of Energy (Award No. DESC0001057 at Cornell). Computation was performed using the Cornell NanoScale Facility, a member of the National Nanotechnology Infrastructure Network, which is supported by the National Science Foundation (Grant ECCS-0335765).